\documentclass[proof]{WileyASNA-v1}

\articletype{Article Type}%

\received{01 November  2022}
\revised{}
\accepted{}

\begin{document}

\title{Detector System Challenges of the Wide-field Spectroscopic Survey Telescope (WST)}

\author[1]{Roland Bacon*}

\author[2,3]{Martin M. Roth}

\author[4]{Paola Amico}

\author[2]{Eloy Hernandez}

\author[5]{The WST consortium}

\authormark{Bacon \textsc{et al}}

\address[1]{\orgdiv{Univ. Lyon 1}, \orgname{CNRS}, \orgaddress{\country{France}}}

\address[2]{\orgdiv{innoFSPEC}, \orgname{Leibniz-Institut f\"ur Astrophysik Potsdam (AIP)}, \orgaddress{\country{Germany}}}

\address[3]{\orgdiv{Institut f\"ur Physik und Astronomie}, \orgname{Universit\"at Potsdam}, \orgaddress{\country{Germany}}}

\address[4]{\orgname{European Southern Observatory}, \orgaddress{\country{Germany}}}

\address[5]{https://www.wstelescope.com}

\corres{*Roland Bacon \email{roland.bacon@univ-lyon1.fr}}

\presentaddress{Univ Lyon, Univ Lyon1, Ens de Lyon, CNRS, Centre de Recherche Astrophysique de Lyon UMR5574, F-69230, Saint-Genis-
Laval, France}

\abstract{The wide-field spectroscopic survey telescope (WST) is proposed to become the next large optical/near infrared facility for the European Southern Observatory (ESO) once the Extremely Large Telescope (ELT) has become operational. While the latter is optimized for unprecedented sensitivity and adaptive-optics assisted image quality over a small field-of-view, WST addresses the need for large survey volumes in spectroscopy with the light-collecting power of a 10\,m class telescope. Its unique layout will feature the combination of multi-object and integral field spectroscopy simultaneously. For the intended capacity of this layout a very large number of detectors is needed. The complexity of the detector systems presents a number of challenges that are discussed with a focus on novel approaches and innovative detector designs that can be expected to emerge over the anticipated 20-year timeline of this project.}



\maketitle


\section{Introduction}\label{sec1}

WST is proposed as an innovative 10-m class wide-field spectroscopic survey telescope with a large field-of-view (5 sq. degree) providing simultaneous operation of a high multiplex (20,000) multi-object spectrograph facility with both medium and high resolution modes (MOS), and a giant panoramic integral field spectrograph (IFS). Through spectroscopic characterization, WST will leverage the understanding of the data from many other huge imaging surveys that merely detect and classify sources. WST will achieve transformative results in most areas of astrophysics (Fig. 1): e.g. the nature and expansion of the dark Universe, the formation of first stars and galaxies and their role in the cosmic reionisation, the study of the dark and baryonic material in the cosmic web, the baryon cycle in galaxies, the formation history of the Milky Way and dwarf galaxies in the Local Group, characterization of exoplanet hosts, and the characterization of transient phenomena. The WST telescope and instruments will be designed as an integrated system and mostly use existing technology, aiming to minimize its carbon footprint and impact on local environment. The ambition is to make WST the next large ESO project after the 39m ELT. ESO is the major inter-governmental organization for European astronomy with a unique suite of large telescopes in Chile. A large consortium, representing 9 European countries and Australia, has formed to start a conceptual design study. In the following, we will present the scientific rationale, the intended layout of the WST, and discuss the specific challenges for the detector subsystems of this complex facility.

\section{Science case }\label{sec2}

\begin{figure*}[th!]
\centerline{\includegraphics[width=12cm]{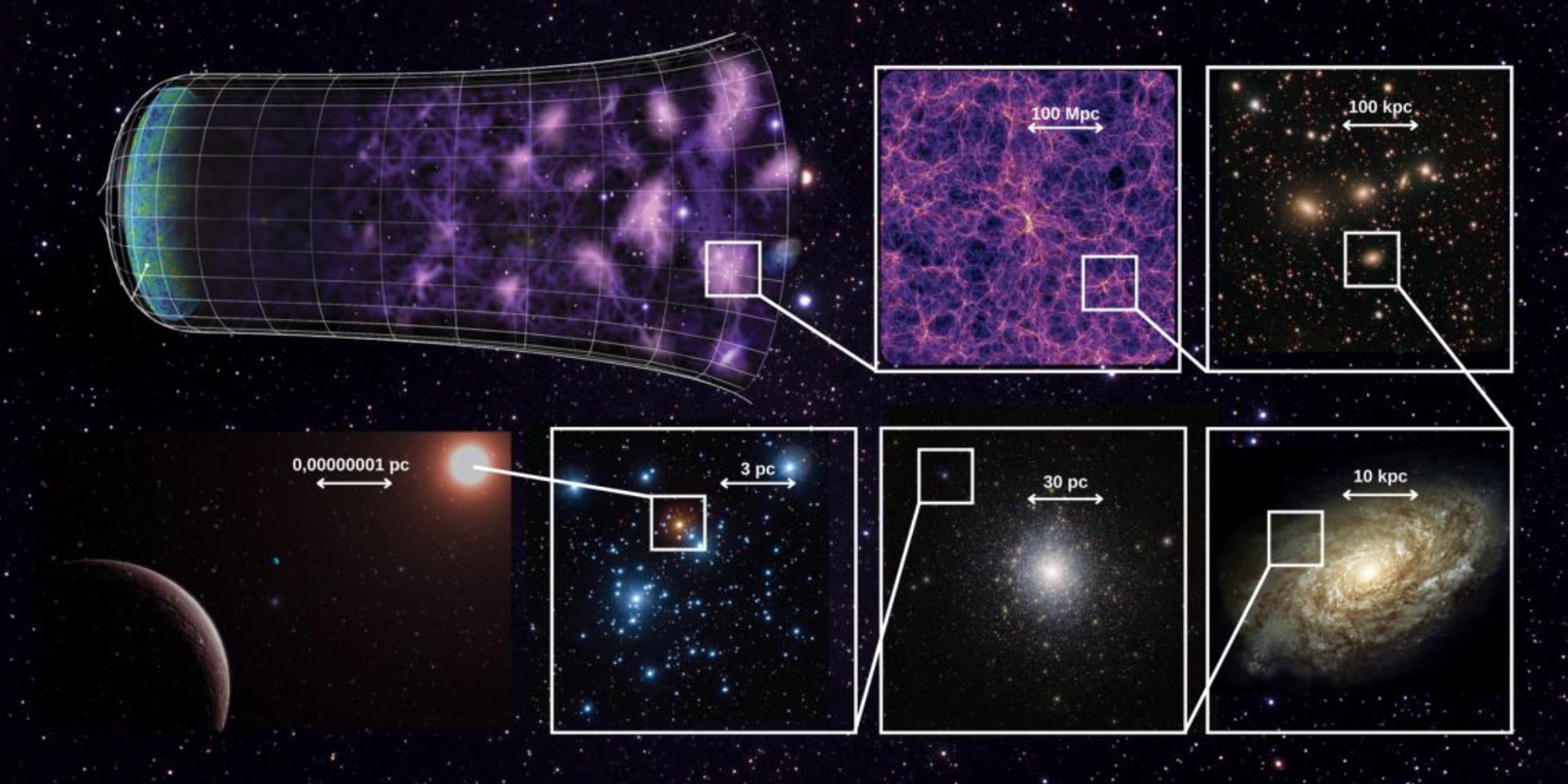}}
\caption{WST will be a revolutionary spectroscopic facility addressing many open questions in astrophysics over a large range in physical scales: from the formation of the large-scale structures in the early universe (100s of Mpc), to the interplay of galaxies in the cosmic web (10s of Mpc), to the formation of our own Galaxy (kpc scales), to the evolution of stars and the formation of planets around them (sub-pc scales).\label{fig:science}}
\end{figure*}

Astrophysics is witnessing a golden era with a number of breakthrough discoveries achieved in the last decade or so, and a large variety of new instrumentation and programs planned for the next decade. Nonetheless, a comprehensive understanding of the formation and evolution of structures in the Universe is still missing. Many key quests in modern astrophysics and cosmology are still in their infancy and require new data to address them:
\begin{itemize}
    \item Is the accelerated expansion of the Universe due to an unknown form of energy or to a modification of General Relativity on large scales? 
    \item What is the interplay between dark, stellar, and gaseous material in galaxies and how does primordial and metal-enriched gas flow in and out of galaxies at various scales?
    \item What is the detailed formation history of our own Galaxy, the Milky Way and of its satellites?
    \item What is the origin of the various chemical elements that are crucial to trace galactic evolution?
    \item What are the conditions that drive the formation and evolution of extra-solar planets?
    \item What are the extreme physical conditions that govern transient events (explosions, eruptions, and disruptions)?
\end{itemize}

A vital element to answering these questions will be the availability of vast all-sky spectroscopic datasets of nearby and distant astrophysical sources at medium and high resolution delivering quantitative measurements of key physical parameters (examples include the redshift of galaxies, z -an indicator of a galaxy distance), the kinematics of galaxies and stars, the physical and evolutionary properties of galaxies, the metal content and detailed chemical composition of stars, etc.).

Over the next decades, we expect a deluge of high-quality data delivered by the upcoming large ground-based (e.g., LSST/VRO, SKA, CTA) and space (e.g., JWST, Euclid, Roman Space Telescope, Athena) telescopes. In parallel the Gaia astrometric mission will have its final release. The imaging capabilities of some of these new facilities will detect and classify a huge number of objects. However, to learn what they are, spectroscopic follow-up on the main representatives as well as rare cases at adequate spectral resolution and cadence is required. Given the expected number of sources (e.g., 20 billion galaxies and 17 billion stars down to R~27.5 for the VRO alone), only a dedicated spectroscopic facility will be able to fully realise the scientific potential of these wide-field imaging surveys. This has been ably demonstrated by the many previous major stellar and extragalactic surveys that have had a huge scientific impact and strong legacy value thanks to their spectroscopic follow-up on photometric and astrometric observations.

Spectroscopic surveys are usually performed with either multi-object spectroscopy (MOS) instruments based on fibre technology (e.g., FLAMES at VLT, \citealt{Pasquini2004}) or multi-slit technology (e.g., VIMOS at VLT, \citealt{Lefevre2003}). New MOS facilities using dedicated 4m-class telescopes are coming on-line (e.g., DESI, 4MOST, WEAVE). In parallel, new MOS instruments will be installed on existing 10m-class multi-purpose telescopes (e.g., PFS at Subaru, MOONS at VLT). These new capabilities will only be able to perform large surveys of bright sources (using the 4m-class facilities full time) or small surveys of faint sources (using part-time access to the 10m-class MOS). The upcoming giant telescopes (ELT, TMT, GMT) are also limited, by technical feasibility, to a small field-of-view and a correspondingly small multiplex capability (e.g., a multiplex of about 300 over 0.12 square degree for MOSAIC at the ELT). Consequently, only a new spectroscopic survey facility on a dedicated wide field-of-view 10m-class telescope equipped with a very high-multiplex MOS will be able to follow-up on the development of upcoming massive multi-wavelength imaging and astrometric capabilities and datasets, as well as conducting efficient follow-up of transients and variable phenomena.

Separately, the advent of panoramic integral-field spectrographs (IFS) on 10m-class telescopes (e.g., MUSE, \citealt{Bacon2010}) has been made possible by well-established IFS expertise in Europe and in particular technological advances in large-format glass slicers. The field of view of the latest IFS, combined with their exquisite sensitivity, is now sufficiently large (e.g., 1 arcmin$^2$ for MUSE) to enable very deep spectroscopic surveys in small patches of the sky. The MOS and IFS approaches to spectroscopy are highly complementary as they probe the sky in very different ways: sparse sampling of sources over a wide area for the MOS and densely packed faint sources in a small area for the IFS. A panoramic IFS also has the unique property of not requiring source pre-selection as needed for the MOS and of providing a 2D continuous sampling of the sky for studies of faint extended spectral signatures (e.g., circumgalactic gas emission). The main limitation of the current panoramic IFS is their relatively small field-of-view. An IFS with a larger-field of view is highly desirable, but given the very large number of spectrographs involved, is not compatible with the existing 10m-class telescope infrastructures. Instead, a new dedicated facility with the right infrastructure is the most appropriate way forward.

The demand for a 10m-class telescope dedicated to spectroscopic surveys is an interest shared worldwide and figures explicitly in many national strategic science plans (e.g., US 2020 decadal survey, 2016-2025 decadal plan for Australian astronomy, Canadian astronomy long range plan 2020-2030). In Europe in particular, a recent poll among the ESO users showed that 75\% of that community identified such a facility as the most crucial one for the future \citep{Merand2021}. The concept of combining both MOS and IFS capabilities in a dedicated wide-field 10m-class spectroscopic telescope would fill a clear gap in the European research infrastructures as described above and ensure full exploitation of the imaging data from upcoming instrumentation. The ambitious WST top-level requirements (Fig. 2) place it far ahead of existing and planned facilities. In just 5 years of operation, the WST MOS will target 250 million galaxies and 25 million stars at medium resolution + 2 million stars at high resolution, and 4 billion spectra with the WST IFS. Such a facility will have a very broad user community, as reflected by the current scientific productivity of VLT instruments. Most importantly, it will enable the scientific community to comprehensively address the questions outlined above and lead to transformative science in a number of key areas, from cosmology to the formation of planets.

\begin{figure}[h]
	\centerline{\includegraphics[width=85mm]{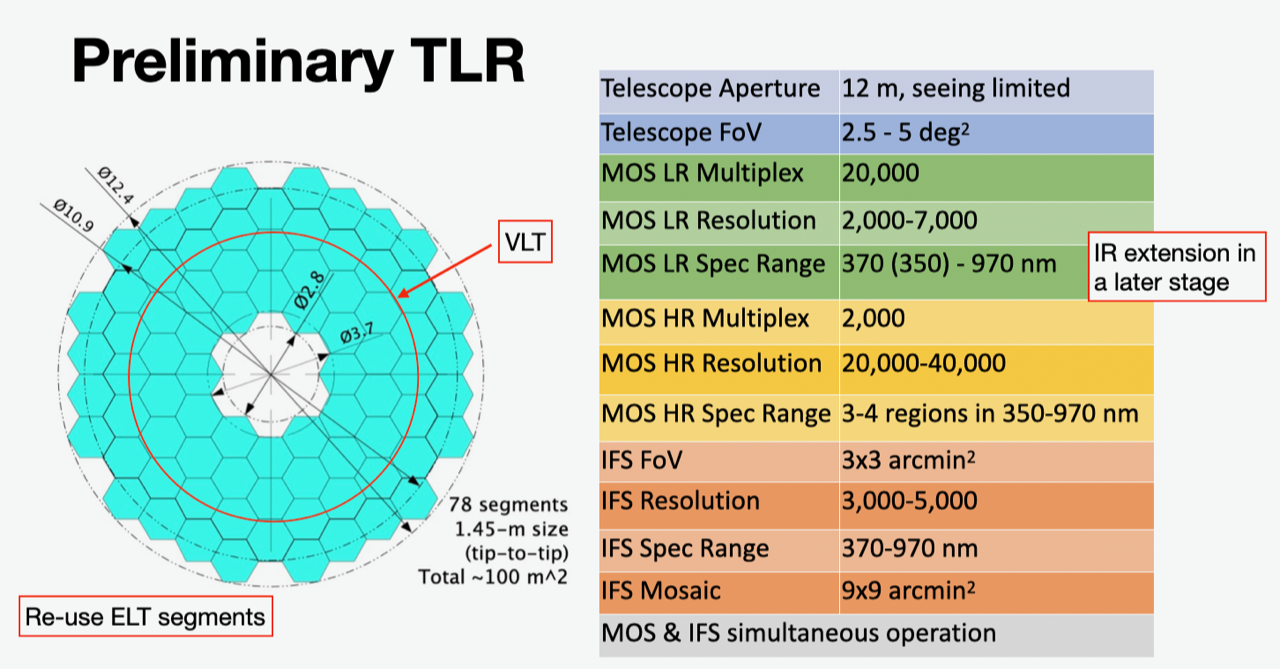}}
	\caption{WST preliminary top level requirements. See also Fig.4\label{fig:prelimTRL}}
\end{figure}

\section{Conceptual Design}\label{sec3}
The consortium has a wealth of experience in the design and construction of similar telescope and instrumentation capabilities. Our initial assessment is that almost all the building blocks are in place. The telescope technologies are available through ELT and other segmented primary 10m-class telescopes worldwide. 

\begin{figure}[h]
	\centerline{\includegraphics[width=78mm]{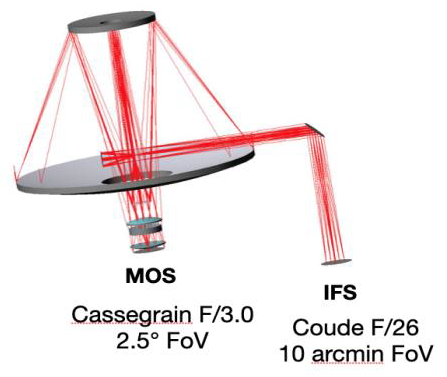}}
	\caption{WST preliminary layout \citep{Pasquini2018}.\label{fig:layout}}
\end{figure}

The partners in this proposal have built a massive panoramic IFS (MUSE) and many fibre MOS instruments and we know we can build individual positioners to the scale needed. We already have a preliminary optical design for the telescope that fulfills the major requirements of aperture and field-of-view (Fig. 3). The main challenge is to move to relatively large-scale production of instrumentation, especially spectrographs, for which the detectors are a critical item. 

\clearpage

\begin{figure}[th!]
	\centerline{\includegraphics[width=85mm]{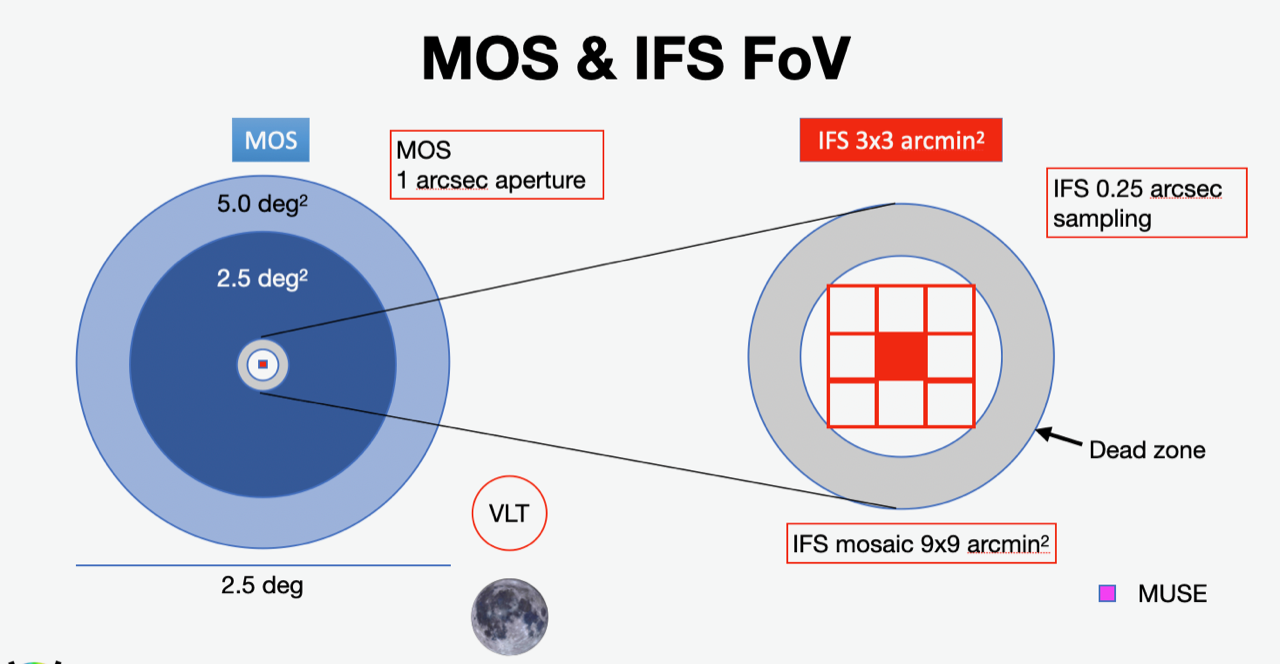}}
	\caption{WST field-of-view.\label{fig:FoV}}
\end{figure}

We will develop close cooperation with industry in the following key areas requiring development to deliver cost effectively in the quantities we need: detectors (see next section), innovative fabrication of image slicers, innovative photopolymer-based volume phase holographic gratings, massive and cost-effective mass production of spectrographs and miniature robots for the positioner. 

\begin{table}[h]
    \centering
    \begin{tabular}{c|r|r|r|l}
    \hline
         &  Spectro. & Chan. & Detec. & Format\\
    \hline
       IFS     & 144 & 2 & 288 & 4k$\times$4k 15 $\mu$m\\
       MOS MR  & 26  & 2 &  52 & 6k$\times$6k 15 $\mu$m\\
       MOS HR  & 4   & 4 &  16 & 9k$\times$9k 10 $\mu$m\\
       Total   & 174 &   & 365 & \\
    \hline
    \end{tabular}
    \caption{Estimated number of spectrographs, channels and detectors (with desired formats) for WST.}
    \label{tab:nbdet}
\end{table}

\section{The WST detector systems}\label{sec4}

The combination of a large aperture telescope with wide field-of-view, massive multiplex, medium to high spectral resolution and large contiguous spectral range is challenging. To maintain the number of spectrographs in an acceptable range, one needs to use spectrographs with a very fast  camera (e.g. f/1). Such spectrographs are usually difficult to build, have a low throughput because of the large number of optical elements, and are very expensive. An elegant solution is to use curved detectors. Curved detectors can greatly simplify the spectrograph camera optics, improving its throughput and reducing its cost \citep{Hugo20191}. 
Curved detectors are currently at a low Technology Readiness Level (TRL). However, a radius of curvature < 200mm for a 4k$\times$4k 15$\mu$m sensor would have very significant cost benefits for the optics. 

\begin{figure}[th!]
	\centerline{
         \includegraphics[width=78mm]{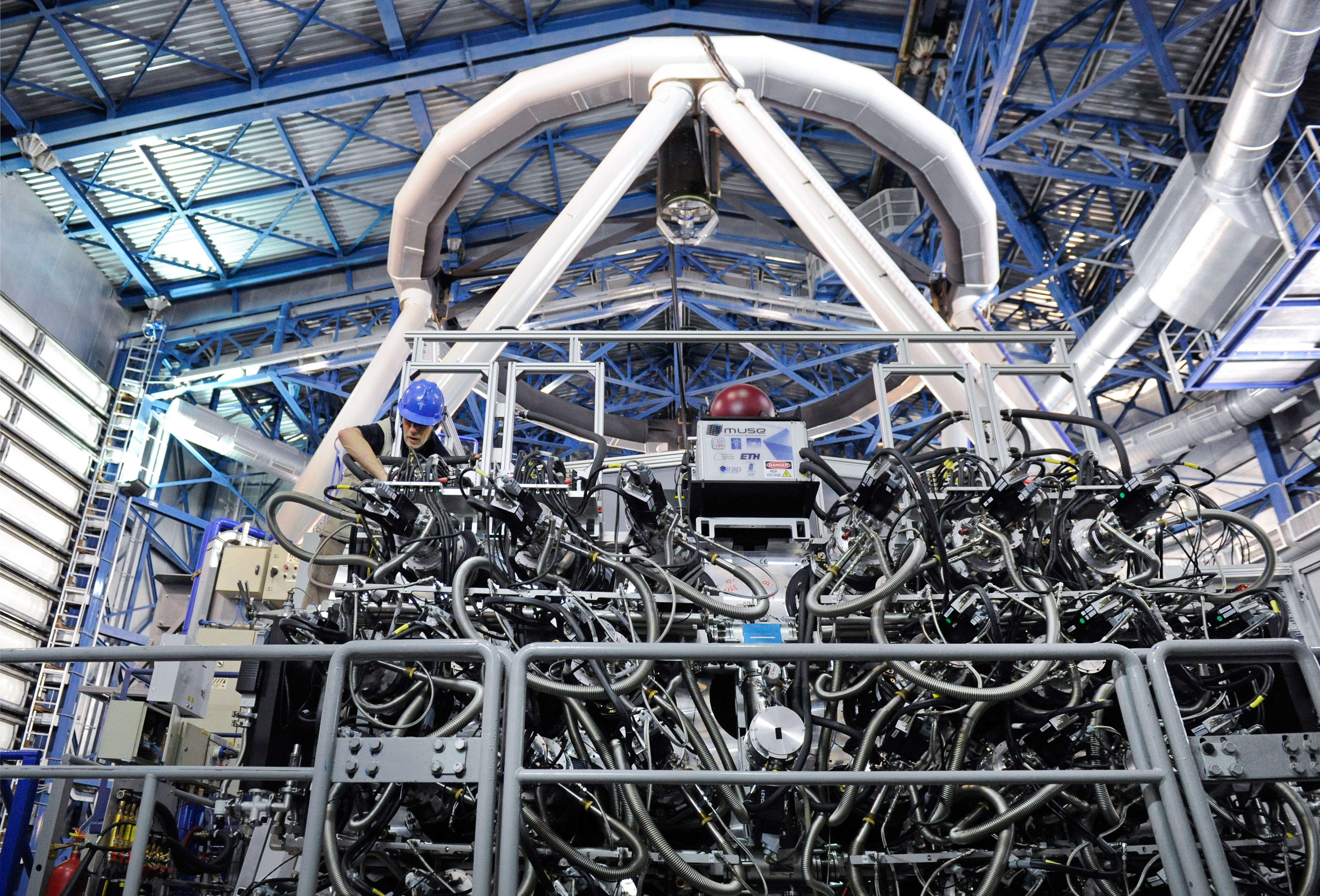}
         }
         \vspace{2mm}
    	\centerline{
         \includegraphics[width=78mm]{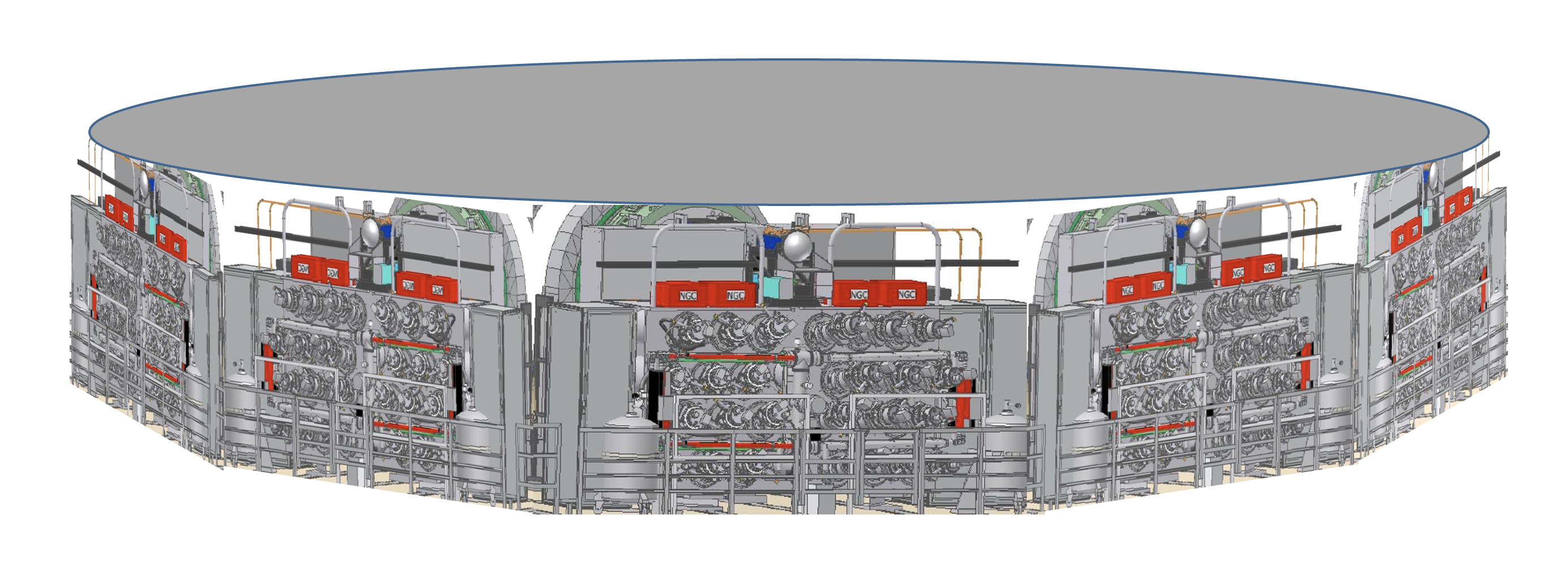}
         }     
	\caption{Top: MUSE at the Nasmyth platform of the VLT. Bottom: artist's view of a possible configuration of a total of 12 MUSE-equivalent spectrograph and detector modules\label{fig:IFU}.}
\end{figure}

Even with fast optics, a large number of spectrographs and an even larger number of detectors is needed (Tab.\ref{tab:nbdet}) to achieve the top level requirements. Given the expected large number of detectors, we will also consider the use of CMOS (whether flat or curved) over traditional CCDs. Thus, we will seek industrial studies to understand the current TRL of these devices and future availability plus costs. In practice, there will be a number of challenging tasks that will be addressed in an upcoming conceptual design study:
\begin{itemize}
    \item Detector selection trade-off: flat versus curved, CMOS versus CCD. Challenges: unproven technologies.
    \item Detector electronics. Challenges: current controller technologies will be outdated, long lead time.
    \item Detector cryogenics. Challenges: complex system for very large number of detectors. LN2 vs. cryo\-cooler. 
    \item Detector procurement and characterization plan. Challenges: massive task, involving industry. 
    \item  Data acquisition system. Challenges: data rates, storage.
    \item  Data preprocessing. Challenges: data rates.
    \item  Thermal management. Challenges: spectrograph thermal stability, potentially high dissipation of heat.
    \item  Mechanical layout and cable routing. Challenges: complexity, interference. 
    \item  Detector QC management. Challenges: complexity, interference.
\end{itemize}

Fig.~\ref{fig:IFU} is an illustration to support the notion that with technologies already available today, WST can indeed be built and operated: the photograph on top shows the existing and highly successful MUSE instrument on the Nasmyth platform of the VLT UT4 that has been in operation at Paranal Observatory since 2014. The rough sketch on the bottom imagines how a total of 12 copies of MUSE modules might look like in a possible configuration for the WST IFU, assuming the number of detectors listed in Tab.~\ref{tab:nbdet}. In fact, the completion of the VIRUS instrument for HETDEX has demonstrated that it is possible to build a complex IFU with as many as 156 detectors \citep{Hill2021}. 

While it is not reasonable to assume that identical copies of a technology that was designed and built between 2003 and 2014 would serve a significantly more complex tasks two decades later, it is entirely plausible to assume that an evolution towards more compact, replicable units is feasible and affordable. Lessons learned with the VIRUS detector system \citep{Hill2016,Hill2021} have prompted us to pay particular attention to technical details that cannot be simply scaled up by number, and that are listed above.

For example, the complex tubing of the vacuum and cryogenic system for MUSE that can be seen in Fig.~\ref{fig:IFU} will almost certainly shrink to a more compact design with new emerging cryocooler technologies that are currently becoming available on the market. 

Interesting trends of CMOS image sensors with detector controller and pre-processor architectures on the same chip may point to revolutionary avenues of creating compact detector modules that would be very different from the bulky CCD detector and controller subsystems of today.

New concepts of distributed data processing units, allocated to individual detectors, similar to ongoing data pre-processing activities for the massive data rates of SKAO \citep{Bonaldi2021}, might significantly reduce data rates, storage capacity, and power consumption - considerations that will play an increasingly important role for green technologies of the future. 

These and other issues will be carefully investigated in an upcoming conceptual design study. The benefits of replicable system modules in terms of cost and reliability, as demonstrated successfully with MUSE, and considered to form the core of the WST layout, have been highlighted by \cite{Hill2014}.

\section{Conclusions}\label{sec5}
The Wide field Spectroscopic Telescope facility is foreseen as a strong candidate for the post-ELT major project at ESO. The detector system will be a major item of WST both in quantity ($\sim$500 sensors) and characteristics (curved sensors). The complexity of this system will likely drive research and develop\-ment concerning detectors and detector electronics, as well as cryogenics, house-keeping and quality control, data acquisition and data processing to new levels, that potentially could result in a paradigm change and a disruptive innovation for detector systems.


\section*{Acknowledgments}
This work was supported by \fundingAgency{BMBF} under Contract No. \fundingNumber{03Z22AN11} and\fundingNumber{03Z22AB1A} .

\nocite{*}
\bibliography{WST-SDW2022}%

\end{document}